\documentclass{aa}
\usepackage{graphicx}
\usepackage{amsmath,amssymb}

\usepackage{txfonts}
\usepackage{natbib}
\usepackage{longtable}
\usepackage{url}

\usepackage[
    colorlinks=true,
    linkcolor=blue,
    citecolor=magenta,
    urlcolor=cyan,
    filecolor=magenta,
    bookmarks=true,
    bookmarksnumbered=true,
    pdfstartview=FitH,
    breaklinks=true,
]{hyperref}

\bibpunct{(}{)}{;}{a}{}{,}

\title{Radio Emission From Low-redshift Little Red Dots}
\titlerunning{Radio census of local LRDs}
\authorrunning{A. Wang et al.}

\offprints{Ailing Wang, wangailing@ynao.ac.cn; Tao An, antao2008@ustc.edu.cn}
\author{Ailing Wang\inst{1}
\and Tao An\inst{2}
\and Ruqiu Lin\inst{3}}

\institute{
Yunnan Observatories, Chinese Academy of Sciences, Kunming 650216, China \email{wangailing@ynao.ac.cn}
\and
Department of Astronomy, University of Science and Technology of China, 96 Jinzhai Road, Hefei, Anhui 230026, China
\email{antao2008@ustc.edu.cn}
\and
Department of Astronomy, University of Massachusetts Amherst, Amherst, MA 01003, USA 
}

\date{Received ; accepted }

\abstract
{High-redshift Little Red Dots (LRDs) are radio faint in existing survey and deep-field samples. Stacking of archival images constrains only the population-mean flux density, leaving the luminosity range and spectral diversity of the class unconstrained. Nearby LRD analogues allow the radio emission and spectra of individual sources to be measured, which is not possible at $z\sim6$.}
{We build a radio census of the reported low-redshift LRDs. We determine their luminosity range and spectral shapes, and  the sensitivity limits set by existing surveys, to provide observational constraints for physical models of the population.}
{Our sample comprises 44 low-redshift LRDs ($0.101 \leq z \leq 0.882$) identified in published studies based on Sloan Digital Sky Survey (SDSS) , Dark Energy Spectroscopic Instrument (DESI) and Large Sky Area Multi-Object Fiber Spectroscopic Telescope (LAMOST) spectroscopy. We measure flux densities at the optical position of each target in LoTSS (144\,MHz), ASKAP (0.9--1.7\,GHz), FIRST/NVSS (1.4\,GHz), and VLASS (3\,GHz). For J1047, we also include the published VLA 5--6\,GHz flux densities.}
{Eight objects are radio-detected: five at 144\,MHz with LoTSS, six at 3\,GHz with VLASS at $\geq3\sigma$, and J1047 through its published VLA data. The rest-frame 144\,MHz luminosities of the LoTSS detections span $1.8\times10^{30}$ to $3.6\times10^{31}\,\mathrm{erg\,s^{-1}\,Hz^{-1}}$. The LoTSS--VLASS spectral indices range from $-0.27\pm0.04$ (J2048) to $-0.78\pm0.11$ (J1200). J1051 lacks 144\,MHz coverage and has $\alpha_{0.9 \rm GHz}^{\rm 3 GHz} \simeq-0.93$. J1047 has a clear deficit at 144 MHz relative to its optically thin GHz spectrum, consistent with a turnover whose position and physical mechanism are not uniquely constrained by the archival data. The higher detection fraction among the Green Pea sample largely reflects their lower redshift: in this subsample LoTSS reaches luminosities roughly an order of magnitude fainter than in the more distant DESI gold sample. If their escaping spectra and radio luminosities are held fixed and extrapolated as power laws, the five LoTSS detections would have observed-frame 144 MHz flux densities of $\sim0.7$--$32\,\mu\mathrm{Jy}$.}
{ Low-redshift LRDs span more than an order of magnitude in radio luminosity among the detections, and more than two orders of magnitude when the J1047 upper limit is included. They also show several distinct spectral shapes, so the optically defined LRD class encompasses diverse radio properties. Current detection statistics are strongly shaped by redshift and survey sensitivity. The source-by-source luminosity limits provide a reference for interpreting future LRD samples.}


\begin{document}
\maketitle

\section{Introduction}
\label{sec:intro}

Little Red Dots (LRDs) are compact, red, broad-line sources discovered in James Webb Space Telescope (JWST) surveys \citep{Larson2023ApJ...953L..29L, 2024ApJ...963..129M,2024ApJ...964...39G,Kocevski2023ApJ...954L...4K, Maiolino2024A&A...691A.145M, Hviding2025A&A...702A..57H, Kocevski2025ApJ...986..126K,de_Graaff2026MNRAS.551g1567D}. Their rest-frame optical spectra often indicate the presence of accreting black holes (BHs) \citep{2024ApJ...963..129M, Rusakov2025}, whereas their weak X-ray and radio emission is atypical of active galactic nuclei (AGN) \citep{Yue2024ApJ...974L..26Y,2025A&A...693L...2P}. Subsequent studies have suggested that their characteristic V-shaped UV-to-optical turnover, together with the more recently identified modified-blackbody-like optical-to-NIR continua, may provide evidence for dense gas surrounding the accreting BHs on sub-parsec scales \citep{Rusakov2025,InayoshiHo2026,Naidu2026Natur.656..329N}.
The presence of dense gas may also affect the observed radio emission. Even if non-thermal synchrotron emission is intrinsically produced, it can be suppressed by the surrounding gas \citep{1997ApJ...485..112B,2021MNRAS.508..680B,2026arXiv260920913T}. Therefore, the weak radio emission observed in LRDs does not necessarily imply the absence of non-thermal activity. Radio observations, particularly the spectral shape, can thus provide constraints on both the intrinsic non-thermal emission and its absorption by the dense nuclear environment.

For high-redshift (high-$z$) LRDs, it is difficult to disentangle intrinsically weak synchrotron emission from suppression by the surrounding nuclear medium. \citet{2025A&A...693L...2P} report 3$\sigma$ mean-stack limits of 8.8 $\mu$Jy in VLASS at 3 GHz and 13.3 $\mu$Jy in FIRST at 1.4 GHz, with the median-stack limits of 10.8 and 17.7 $\mu$Jy, respectively. Deeper observations constrain a different selection: the GOODS-N stack of 37 JWST-selected, spectroscopically confirmed broad-line AGN reaches an rms of 0.15 $\mu$Jy at 3 GHz \citep{2026A&A...706A.372M}. Whereas this is a broad-line AGN sample, rather than a uniformly selected LRD sample. These results demonstrate faint average radio emission in the selected samples, while leaving the individual luminosity distribution and the incidence of visible radio components uncertain. 

Low-redshift (Low-$z$, $z<1$) LRDs (LLRDs) may provide an opportunity to resolve this ambiguity. Owing to their proximity, their radio emission may be individually detectable through targeted VLA follow-up observations, and Very Long Baseline Interferometry (VLBI) observations can further probe compact radio emission from their nuclear regions.

Several such populations have now been identified spectroscopically. They include seven V-shaped broad-line Green Peas (BLGP) \citep{2025ApJ...980L..34L}, three nearby Sloan Digital Sky Survey (SDSS) LRDs \citep{Ji2026MNRAS.545f2235J,2026ApJ...997..364L,2026A&A...707L..17R}, a larger Dark Energy Spectroscopic Instrument (DESI) DR1 population \citep{2026arXiv260521574L,2026arXiv260514233P,2026A&A...709L..11D}, and an infrared-luminous galaxy J2048 \citep{2026ApJ...999...30C,2026arXiv260816200R}. The searches applied different optical criteria, and together they define the published set of low-redshift LRDs whose radio properties can presently be examined source by source.

The existing detections of LLRDs already point to a range of radio behaviour. J1047 contains a compact and stable 6 GHz source with an optically thin in-band spectrum \citep{2026A&A...707L..17R}. J2048 is substantially brighter across the centimetre band \citep{2026ApJ...999...30C,2026arXiv260816200R}. J0923 is detected at 144 MHz at the optical position (An et al. 2026). These measurements come from different surveys, angular resolutions, detection procedures and frequency ranges, and catalogue detections alone discard the information carried by objects that fall below a formal source-extraction threshold. Establishing the luminosity range, the spectral diversity and the survey-selection boundaries of the population therefore calls for one common set of measurements.

We construct a 44-object working sample comprising literature LLRDs and LRD analogues as well. We measure the radio flux density at the optical position from 144 MHz to 3 GHz using the LOFAR Two-metre Sky Survey (LoTSS; \citealt{2022A&A...659A...1S}), the Australian Square Kilometre Array Pathfinder surveys RACS (Rapid ASKAP Continuum Survey) and VAST (Variables and Slow Transients; ASKAP; \citealt{2020PASA...37...48M,2021PASA...38...58H,2021PASA...38...54M}), the Faint Images of the Radio Sky at Twenty Centimeters survey (FIRST; \citealt{1995ApJ...450..559B}), the NRAO (National Radio Astronomy Observatory) VLA Sky Survey (NVSS; \citealt{1998AJ....115.1693C}), and the Very Large Array Sky Survey (VLASS; \citealt{2020PASP..132c5001L}), and, as auxiliary measurements, the low-frequency TIFR GMRT Sky Survey \citep[TGSS;][]{2017A&A...598A..78I} and the GaLactic and Extragalactic All-sky MWA Survey \citep[GLEAM;][]{2017MNRAS.464.1146H}. We also incorporate the published 5--6 GHz flux densities of J1047 \citep{2026A&A...707L..17R}.

The first paper of this series (An et al. 2026a) presented J0923 and, through comparison with J1047 and J2048, drew attention to the diversity of radio properties among local LRDs. Here, this paper extends that source-level comparison to the full published LLRD sample and places the diversity on a common observational basis. Three empirical questions organise the analysis. What range of radio luminosities does the present survey material reveal? Do the radio-detected sources occupy a common spectral state? Which population-level trends survive the strongly redshift-dependent survey limits? Paper~III of the series (An et al. 2026b) will combine the radio measurements with optical and gas diagnostics and address the physical origin of the diversity\footnote{Throughout, we use the definition $S_\nu\propto\nu^\alpha$. Monochromatic luminosities are calculated as $L_\nu = 4\pi D_{\rm L}^{2}S_\nu(1+z)^{-1-\alpha}$ for $H_0=70~{\rm km\,s^{-1}\,Mpc^{-1}}$ and $\Omega_{\rm m}=0.3$.}.

\section{Sample}
\label{sec:sample}

\begin{figure}
\centering
\includegraphics[width=\hsize]{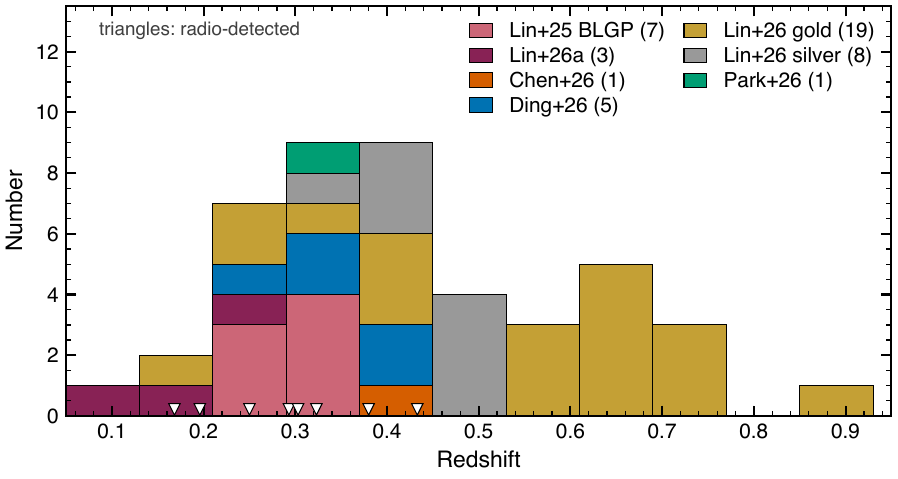}
\caption{Redshift distribution of the 44 LLRDs, stacked by parent catalogue. Filled triangles mark the eight radio-detected sources. The Green Peas of \citet{2025ApJ...980L..34L} sit near $z\sim0.3$, and the DESI gold sample of \citet{2026arXiv260521574L}, which dominates the count, near $z\sim0.6$.}
\label{fig:z}
\end{figure}

Our catalogue comprises 44 distinct sources spanning $0.101\le z\le0.882$ (Table~\ref{tab:subs} and Fig.~\ref{fig:z}). It combines objects identified in eight published searches with different selection functions, and therefore does not constitute a uniformly selected or complete sample.
We retain the original parent-sample classifications to preserve the each source's selection history. Radio detections are therefore interpreted in the context of the corresponding parent sample rather than combined into a single population-wide detection fraction.

\begin{table}
\caption{Parent samples and radio coverage. $N^{\rm LoTSS}_{144 \rm MHz}$ is the number of targets with a finite LoTSS pixel at the optical position, nine of the 44 falling in mosaic gaps or outside the survey footprint. $N^{\rm VLASS}_{3 \rm GHz}$ is the number with at least one VLASS Quick Look image, and all 44 lie in the VLASS footprint. $N_{\rm det}$ is the number satisfying the detection criteria of Sect.~\ref{sec:assoc}. Seven sources from \citet{2026arXiv260514233P} overlap with the sample of \citet{2026arXiv260521574L} and are counted only once.}
\label{tab:subs}
\centering
\small
\begin{tabular}{lrrrrr}
\hline\hline
Parent & $N$ & $z_{\rm med}$ & $N^{\rm LoTSS}_{144 \rm MHz}$ & $N^{\rm VLASS}_{3 \rm GHz}$ & $N_{\rm det}$ \\
\hline
\citet{2025ApJ...980L..34L} BLGP     &  7 & 0.293 & 7  &  7 & 3 \\
\citet{2026ApJ...997..364L} SDSS     &  3 & 0.168 & 3  &  3 & 1 \\
\citet{2026ApJ...999...30C} J2048    &  1 & 0.433 & 1  &  1 & 1 \\
Ding et al. (2026)         &  5 & 0.340 & 1  &  5 & 2 \\
\citet{2026arXiv260521574L} gold     & 19 & 0.608 & 14 & 19 & 1 \\
\citet{2026arXiv260521574L} silver   &  8 & 0.447 & 8  &  8 & 0 \\
\citet{2026arXiv260514233P}          &  1 & 0.291 & 1  &  1 & 0 \\
\hline
Total            & 44 & 0.399 & 35 & 44 & 8 \\
\hline
\end{tabular}
\end{table}

The seven objects from \citet{2025ApJ...980L..34L} are the V-shaped sources of their broader broad-line Green Pea sample. The twelve broad-line Green Peas without V-shaped continua are not included here. \citet{2026ApJ...997..364L} and \cite{Ji2026MNRAS.545f2235J} contribute three nearby spectroscopic analogues, J1022, J1025, and J1047, which are the only subsample with published high-resolution VLA imaging. The DESI DR1 search of \citet{2026arXiv260521574L} provides 27 objects, classified in the parent study as 19 gold and eight silver LRDs, of which 18 are reported to show Balmer absorption. \citet{2026A&A...709L..11D} add five DESI-selected analogues, four at $\delta<0^\circ$. Seven objects in \citet{2026arXiv260514233P} overlap with the \citet{2026arXiv260521574L} sample, leaving J0716+5433 as the only additional source. J2048 is treated separately because it was identified as an ultra-luminous infrared galaxy (ULIRG)-like LRD analogue with established multi-band radio emission \citep{2026ApJ...999...30C}.

We retain the parent-study labels, including Green Pea, blackbody (BB)-like, Balmer absorption, and gold/silver classification, to document selection history (Table \ref{tab:subs}). Optical results are adopted from those studies, and a homogeneous optical characterisation is deferred to Paper III. X-ray weakness is assigned only where usable X-ray observations support it. Absence of coverage is not equivalent to a non-detection.

\section{Radio data and measurements}
\label{sec:data}

\subsection{Surveys}
\label{sec:surveys}

LoTSS provides the 144 MHz measurements \citep{2022A&A...659A...1S}. Usable mosaic data are available at the optical positions of 35 of the 44 targets. Two sources are covered by DR2 and 33 by DR3. The restoring beam is approximately $6''$ for the DR2 fields and the northern DR3 coverage, increasing to approximately $9''$ at lower declinations. The remaining nine targets have no usable LoTSS measurement. Four fall in mosaic gaps, two sources from \citet{2026A&A...709L..11D} lie south of the survey footprint, and three northern sources, including J165450+033741, lack usable local imaging. These objects are treated as having no 144 MHz coverage rather than as radio non-detections.

ASKAP observations from RACS and VAST \citep{2020PASA...37...48M,2021PASA...38...58H,2021PASA...38...54M} provide complementary coverage between 0.84 and 1.66 GHz for 11 targets. After retaining Stokes $I$ images with a complete fitting region, the data set comprises 143 images, corresponding to 99 independent source--epoch--frequency measurements. The restoring beams span approximately $9''$--$25''$, and the observing cadence, frequency coverage, and angular resolution vary between products. The ASKAP measurements therefore serve to constrain spectral shape and multi-year persistence for the sources with suitable coverage, and they stay outside the census itself.

VLASS provides the uniform 3 GHz coverage for the full sample \citep{2020PASP..132c5001L}. We use the Quick Look images from epochs 1--3 spanning 2017 September to 2024 July. The data set contains 131 independent Quick Look images. The restoring beams range from approximately $2.3''\times1.8''$ to $5.0''\times2.7''$. The VLASS Quick Look products are subject to campaign-dependent peak-flux-density biases at the $\sim10$--15\% level and to residual astrometric uncertainties \citep{NRAO2026}. Because these effects vary between campaigns, we retain the measured image-plane amplitudes and treat the flux-density scale as a systematic uncertainty separate from the local map noise used for the statistical errors.

FIRST and NVSS provide additional measurements near 1.4 GHz \citep{1995ApJ...450..559B,1998AJ....115.1693C}. The $5.4''$ resolution of FIRST is sufficiently close to that of LoTSS and VLASS to provide a useful intermediate-frequency constraint for compact sources. The $45''$ NVSS beam is substantially more susceptible to confusion and extended emission, so the NVSS measurements are retained as ancillary flux-density information but are excluded from the spectral-index analysis.

For J1047, we also adopt the published VLA measurements of \citet{2026A&A...707L..17R}: $130\pm9 \,\mu\mathrm{Jy}$ at 5 GHz from the 2010 C-configuration observation and $117\pm8 \,\mu\mathrm{Jy}$ at 6 GHz from the 2018 A-configuration observation. The latter has an in-band spectral index of $\alpha=-0.85\pm0.24$. After accounting for the frequency difference using this spectral index, the two epochs show no significant variability over the eight-year interval.

\subsection{Flux density measurements}
\label{sec:phot}

All survey images are analysed with optical-position forced photometry following An et al. (2026). At each target position, we fit a unit-peak elliptical Gaussian fixed to the restoring beam in major axis, minor axis, and position angle, with its centroid fixed at the adopted optical coordinates and a local background term. The fitted amplitude gives the total flux density of an unresolved source. For resolved emission, it measures the beam-scale component centred on the optical position, which need not equal the integrated flux density (Sect.~\ref{sec:res}). The fitted amplitude is allowed to be either positive or negative, ensuring an unbiased measurement in the presence of noise, particularly for non-detections. Fixing the position a prior avoids the positive bias introduced by selecting nearby noise peaks or unrelated radio sources.

For a baseline fit, we use a circular region of radius $1.6\,\theta_{\rm maj}$, where $\theta_{\rm maj}$ is the FWHM of the restoring-beam major axis. We then repeat the fit with radii of $1.2$, $1.6$, and $2.0\,\theta_{\rm maj}$ for each sources with constant and planar backgrounds, giving six variants.
The statistical uncertainty, $\sigma_{\rm stat}$, is taken as the largest of three empirical estimates: the local background rms; the robust scatter of identical fixed--beam Gaussian fits at blank positions sampled at $\approx1.4\,\theta_{\rm maj}$ spacing across the cutout, excluding positions within $2.5\,\theta_{\rm maj}$ of the target and using at most 120 positions per image; and the scatter among the six fitting variants. We adopt the largest of the three rather than combining them in quadrature, since the estimates draw on the same correlated image pixels and are not statistically independent.

For LoTSS, we consider the forced photometry as the primary measurement, and we perform an independent free-centroid fit with the same beam template as an astrometric check for Sect.~\ref{sec:assoc}. 
For LoTSS, the quoted flux-density uncertainties are the statistical uncertainties, $\sigma_{\rm stat}$, estimated from our forced-photometry procedure (Table \ref{tab:eight} and Table \ref{tab:full}). 

We apply the same flux-measurement method to VLASS, ASKAP, FIRST and NVSS using the restoring beam of each image. Individual epochs are measured at their native resolution. Multiple-epoch observations within the same survey band are combined as an inverse-variance weighted mean, $\bar{S}_\nu=\sum_i S_{\nu,i}\sigma_i^{-2}/\sum_i\sigma_i^{-2}$, with $\sigma_{\bar{S}}=(\sum_i\sigma_i^{-2})^{-1/2}$. The weights contain only the empirical statistical errors. VLASS image-plane co-additions are used for visualisation only. Tabulated flux densities are obtained from the individual Quick Look images before combination.

Every valid image contributes an amplitude and its uncertainty, $S_\nu\pm\sigma_{\rm stat}$, whatever the significance, and negative amplitudes are kept as measured. This preserves the noise distribution about zero and the information carried by low-significance excursions of either sign. Where a figure or a comparison with earlier work calls for an upper limit, we quote the derived $3\sigma_{\rm stat}$ value and use the underlying amplitude in every calculation.

\subsection{Detection criteria}
\label{sec:assoc}

We define a detection based on LoTSS at 144\,MHz and VLASS at 3\,GHz together with published VLA detections, as these are the only surveys with near-uniform coverage of the full sample. Because every amplitude is measured at a fixed position, we adopt $S_\nu/\sigma_{\rm stat}\geq3$ as the detection threshold. Any detections with ASKAP and FIRST is used for constructing SEDs and variability.

We classify a source as radio detected only when significant radio emission is positionally associated with the optical nucleus. Nearby/off-nuclear radio sources and low-significance signals appearing only in individual epochs are not considered detections.
More specifically, for LoTSS, detection requires $S_{144}/\sigma_{\rm stat}\geq3$ at the optical position together with a free-centroid offset within half a beam FWHM, and the forced-photometry amplitude is the reported flux density. 
For VLASS, radio detections are classified based on the signal-to-noise ratio of the inverse-variance weighted mean flux density across epochs, $\bar{S}_{3}/\sigma_{\bar{S}}$, rather than that of any individual epoch. Individual-epoch measurements are used to investigate radio variability, but variability itself is not used as a detection criterion.
A published VLA detection is also accepted when the radio source is coincident with the optical nucleus within the astrometric precision of the observation. Eight objects satisfy these criteria. Five are LoTSS detections and six are VLASS detections ($\geq 3\sigma$), the two sets sharing J0923, J1200, J2048 and J2146, and J1047 enters through its published VLA core. 

\subsection{Angular resolution and counterpart validation}
\label{sec:res}

Given the limited angular resolution and the different angular resolutions of the radio surveys, the detected emission is positionally associated with LLRDs at the angular resolution for the corresponding survey. To identify possible nearby contamination, we inspect the morphology and free-centroid position for each source. 

Resolution differences also limit spectral comparisons: a larger beam captures extended host emission resolved out at higher resolution, so a lower-resolution flux density exceeding the VLASS value does not necessarily imply spectral curvature. Spectral indices are therefore derived from LoTSS, ASKAP, FIRST, and VLASS, which have relatively high angular resolutions among the available surveys. NVSS, TGSS, and GLEAM are excluded because their substantially larger beams can mix nuclear, host-galaxy, and unrelated extended emission. The LoTSS--VLASS spectral index $\alpha_{144}^{3000}$ is an empirical descriptor of the spectrum between the two bands and carries no assumption that the same physical component dominates at both frequencies. Cases where lower-resolution flux densities exceed the VLASS value, as in J2048 and J2146, are treated source by source.

\section{Results}
\label{sec:results}
 
\subsection{Coverage, depth, and threshold stability}
\label{sec:depth}
 
Among the 35 targets with valid LoTSS data, the median statistical uncertainty is $0.096\,\mathrm{mJy}$, and the median uncertainty of the 30 LoTSS non-detections is $0.088\,\mathrm{mJy}$. Among the 44 VLASS inverse-variance means it is $0.086\,\mathrm{mJy}$. Five targets reach $\mathrm{S/N}\ge5$ at 144\,MHz and four at 3\,GHz.
 
At 144\,MHz no target falls between $3\sigma$ and $5\sigma$. At 3\,GHz the interval holds J1717 at $4.0\sigma$ and J1200 at $3.0\sigma$, and J1200 is already a $16\sigma$ LoTSS source. Varying the detection threshold between $3\sigma$ and $5\sigma$ affects only the classification of J1717, while leaving the luminosity range, spectral classifications, and population-level conclusions unchanged. The measured amplitudes show a clear separation between a noise-dominated majority and a small number of high-significance detections, with few sources at intermediate significance.

\begin{figure*}
\centering
\includegraphics[width=0.9\textwidth]{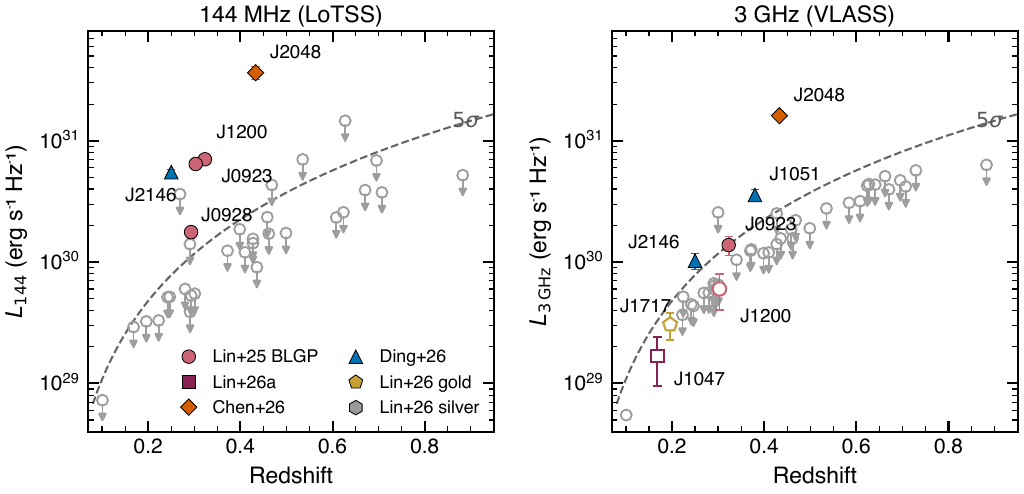}
\caption{Rest-frame luminosity against redshift at 144\,MHz (left) and 3\,GHz (right), computed throughout with $\alpha=-0.7$ so that the detections and the sensitivity curves share one scale. Filled symbols are measurements at $\mathrm{S/N}\ge5$, open squares are VLASS means between $2\sigma$ and $5\sigma$ (J1047, J1717, J1200), and grey symbols are $3\sigma$ limits. The dashed line is the $5\sigma$ point-source depth for the median map noise. }
\label{fig:lz}
\end{figure*}

The wide redshift range of our sample ($0.101\leq z\leq0.882$) leads to substantially different radio luminosity limits. Assuming $\alpha=-0.7$, the nominal $5\sigma$ luminosity limits span a factor of 126 (Fig.~\ref{fig:lz}). Consequently, high-redshift non-detections provide much weaker constraints on intrinsic radio luminosity than low-redshift non-detections. The limiting radio luminosity is therefore not uniform across the sample.

The five LoTSS detections have median $z=0.303$, and the 30 LoTSS non-detections have median $z=0.419$. The median $3\sigma$ flux density limit among the non-detections is $0.26\,\mathrm{mJy}$, corresponding to $L_{144}\simeq1.5\times10^{30}\,\mathrm{erg\,s^{-1}\,Hz^{-1}}$ at $z=0.419$. J0928, the faintest LoTSS detection at $L_{144}=1.8\times10^{30}\,\mathrm{erg\,s^{-1}\,Hz^{-1}}$, would deliver $0.13\,\mathrm{mJy}$ at the DESI gold-sample median redshift $z=0.608$, which is a $1.5\sigma$ measurement in a median LoTSS image. The detection fraction therefore depends strongly on redshift.

\subsection{Individual Radio-detected Sources}
\label{sec:eight}

\begin{table*}
\caption{Radio-detected local LRDs. Flux densities are forced-photometry amplitudes at the optical position, and the VLASS and ASKAP entries are inverse-variance means over epochs (Sect.~\ref{sec:phot}). $\alpha_{144}^{3000}$ is the two-point spectral index between the LoTSS and VLASS measurements. J1047 shows a low-frequency suppression with an uncertain turnover, and J1717 has no useful two-point spectral index. J1051 has no LoTSS coverage, and its ASKAP--VLASS spectral index is $\alpha_{0.9}^{3}\simeq-0.93$. $*$ denote measurements below the nominal $3\sigma$ threshold.}
\label{tab:eight}
\centering
\small
\begin{tabular}{llcllllc}
\hline\hline
Name & Parent & $z$ & $S^{\rm LoTSS}_{144 \rm MHz}$ & $S^{\rm ASKAP}_{0.9 \rm GHz}$ & $S^{\rm FIRST}_{1.4 \rm GHz}$ & $S^{\rm VLASS}_{3\rm GHz}$ & $\alpha_{144}^{3000}$ \\
     &        &     & (mJy)     & (mJy)     & (mJy)     & (mJy)   &  \\
\hline
J0923 & \citet{2025ApJ...980L..34L} BLGP & 0.323 & $2.23\pm0.10$ & \ldots & $0.42\pm0.14$ & $0.44\pm0.08$ & $-0.54\pm0.06$ \\
J0928 & \citet{2025ApJ...980L..34L} BLGP & 0.293 & $0.70\pm0.08$ & $0.40\pm0.14^{*}$ & $0.15\pm0.15^{*}$ & $0.07\pm0.08^{*}$ & $-0.76\pm0.40^{*}$ \\
J1047 & \citet{2026ApJ...997..364L}      & 0.168 & $0.07\pm0.13^{*}$ & $0.32\pm0.33^{*}$ & $0.04\pm0.14^{*}$ & $0.23\pm0.10^{*}$ & turnover \\
J1051 & Ding et al. (2026)      & 0.380 & \ldots        & $2.45\pm0.07$ & $1.23\pm0.16$ & $0.79\pm0.09$ & \ldots \\
J1200 & \citet{2025ApJ...980L..34L} BLGP & 0.303 & $2.36\pm0.14$ & $1.23\pm0.16$ & $0.59\pm0.18$ & $0.22\pm0.07$ & $-0.78\pm0.11$ \\
J1717 & \citet{2026arXiv260521574L} gold  & 0.196 & $0.20\pm0.10^{*}$ & $0.10\pm0.11^{*}$ & $0.11\pm0.14^{*}$ & $0.29\pm0.07$ & \ldots \\
J2048 & \citet{2026ApJ...999...30C}      & 0.433 & $5.94\pm0.75$ & $4.70\pm0.04$ & $3.61\pm0.16$ & $2.63\pm0.09$ & $-0.27\pm0.04$ \\
J2146 & Ding et al. (2026)       & 0.250 & $3.10\pm0.18$ & $2.19\pm0.10$ & $0.84\pm0.20$ & $0.58\pm0.09$ & $-0.55\pm0.05$ \\
\hline
\end{tabular}
\end{table*}

\begin{figure*}
\centering
\includegraphics[width=0.9\textwidth]{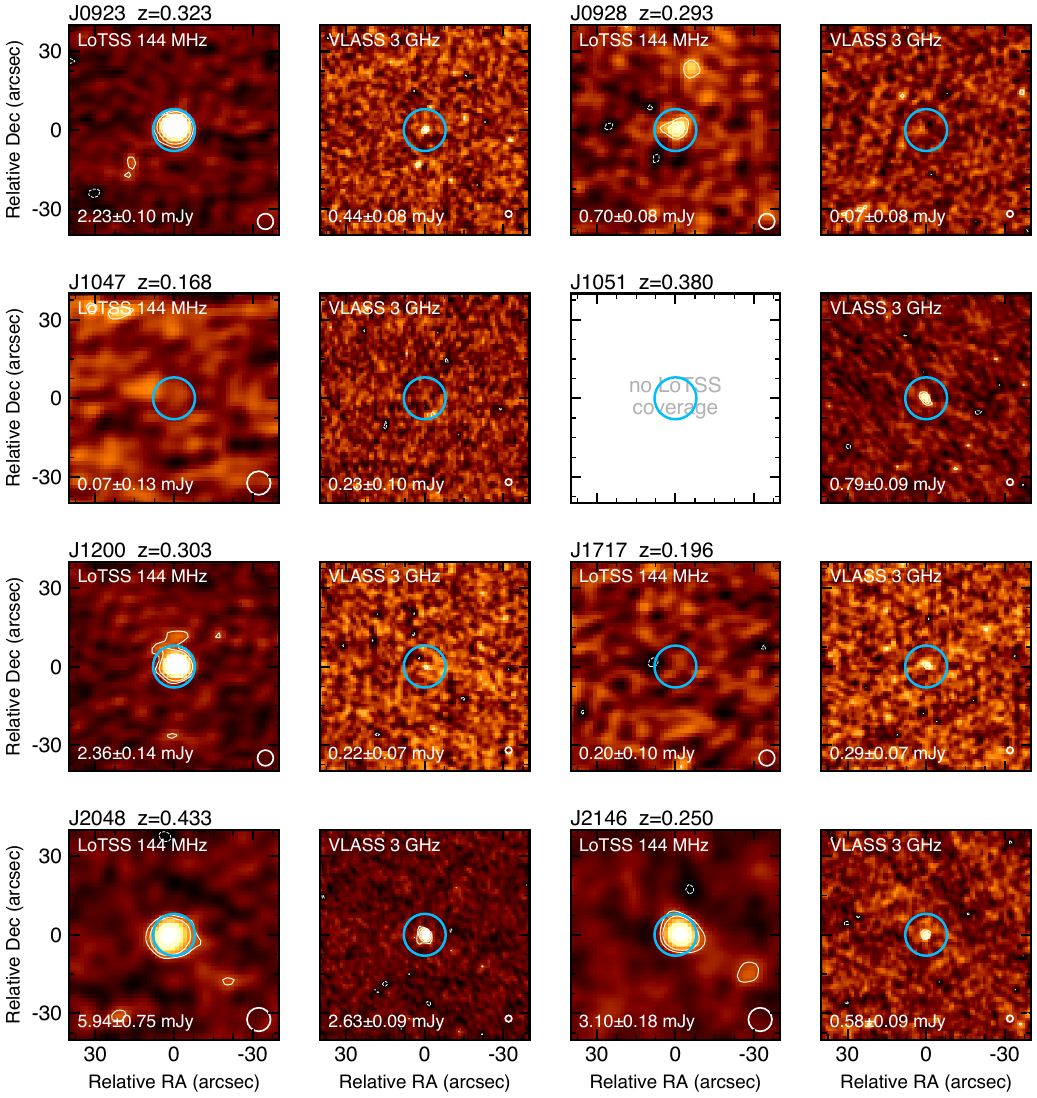}
\caption{Optical-position stamps of the eight radio-detected LLRDs. Each pair shows the LoTSS DR2/DR3 144\,MHz mosaic (left) and the VLASS 3\,GHz co-addition of three epochs (right). Panels span $80''\times80''$ and the greyscale gives the brightness. The ellipse in the lower right corner is the restoring beam. Solid contours are drawn at 3, 5, 10 and $20\sigma$ where present. J1051 has no local LoTSS image.}
\label{fig:stamps}
\end{figure*}

The eight detected targets are shown in Table~\ref{tab:eight} and  Fig.~\ref{fig:stamps}.

\paragraph{J0923, J1200 and J0928.}
These three Green Peas are the 144\,MHz detections at $z\simeq0.3$. J0923 is labeled as BB-like and X-ray-weak. 
LoTSS DR3 gives $S_{144}=2.23\pm0.10\, \mathrm{mJy}$ ($21\sigma$) at the optical position, with a free-centroid offset of only $0.25''$ on a $6''$ beam (Paper I). 
Its VLASS measurements are $0.36\pm0.14$, $0.26\pm0.13$ and $0.75\pm0.17\,\mathrm{mJy}$ in three epochs, and an inverse-variance mean is $0.44\pm0.08\,\mathrm{mJy}$. J1200 is comparably bright with $S_{144} =2.36\pm0.14\,\mathrm{mJy}$ and sits at the $3\sigma$ boundary at 3\,GHz. J0928 is three times fainter at 144\,MHz ($0.70\pm0.08\,\mathrm{mJy}$) and is undetected by VLASS ($0.07\pm0.08\,\mathrm{mJy}$).

\paragraph{J2146.}
DESI J214655+102813 at $z=0.25$ is detected in both 144 MHz and 3 GHz. LoTSS gives $S_{144}=3.10\pm0.18\,\mathrm{mJy}$ on a $9''$ beam and VLASS gives $0.58\pm0.09\,\mathrm{mJy}$, with a free-centroid offset of $0.78''$. A $27.9\,\mathrm{mJy}$ LoTSS source lies $68''$ from the optical position, and it enters the NVSS beam, so the NVSS amplitude is discarded.

\paragraph{J2048.}
J204837.26$-$002437 is the brightest object in every band we measure. LoTSS gives $S_{144}=5.94\pm0.75\,\mathrm{mJy}$, ASKAP-low $4.70\pm0.04\,\mathrm{mJy}$ over 33 epochs, FIRST $3.61\pm0.16\,\mathrm{mJy}$ and VLASS $2.63\pm0.09\,\mathrm{mJy}$. The LoTSS free-centroid offset is $2.55''$ on a $9''$ beam and the free-centroid flux density, $6.61\,\mathrm{mJy}$, exceeds the forced value, so a modest extended contribution is possible at 144\,MHz. The three VLASS centroids lie within $0.3''$ of the optical position. J2048 is estimated with a star-formation rate of several hundred $\mathrm{M}_\odot\,\mathrm{yr}^{-1}$ \citep{2026ApJ...999...30C}, and its 144\,MHz luminosity, $L_{144}=3.6\times10^{31}\,\mathrm{erg\,s^{-1}\,Hz^{-1}}$ for $\alpha=-0.7$, is of the order expected from that rate \citep{2021A&A...648A...6S}. The radio luminosity is thus energetically consistent with star formation; however, the spectral shape (Sect.~\ref{sec:spec}) is not.

\paragraph{J1051.}
DESI J105157$-$062625 lies outside the LoTSS coverage but is a secure VLASS source, $0.79\pm0.09\,\mathrm{mJy}$ in the multi-epoch mean and $3.3\sigma$, $5.7\sigma$ and $5.5\sigma$ in individual epochs. ASKAP-low recovers it at $2.45\pm0.07\,\mathrm{mJy}$ and FIRST at $1.23\pm0.16\,\mathrm{mJy}$, which gives a steep spectral index between 0.9 and 3\,GHz, $\alpha_{0.9}^{3}\simeq-0.93$. 
 
\paragraph{J1047.}
J104755.92+073951 ($z=0.168$) carries a secure GHz core and no metre-wave counterpart. At the optical position LoTSS gives $0.07\pm0.13\,\mathrm{mJy}$, corresponding to a $3 \sigma$ upper limit of $0.39\,\mathrm{mJy}$. The published 6\,GHz source is $117\pm8\,\mu\mathrm{Jy}$ with an optically thin in-band spectral index $\alpha=-0.85\pm0.24$ \citep{2026A&A...707L..17R}, and that power law predicts $2.79\,\mathrm{mJy}$ at 144\,MHz, with a $1\sigma$ range of $1.14$--$6.82\,\mathrm{mJy}$ from the uncertainty on $\alpha$ alone. The LoTSS limit lies a factor of seven below the point estimate and a factor of three below the flattest $1\sigma$ extrapolation. The same power law predicts $0.21\,\mathrm{mJy}$ at 3\,GHz, and the VLASS mean is $0.23\pm0.10\,\mathrm{mJy}$. 

The LoTSS non-detection is difficult to reconcile with an unbroken optically thin spectrum extrapolated from $6\,\mathrm{GHz}$. Differences in angular resolution alone are unlikely to explain the LoTSS non-detection, since the $9''$ LoTSS beam encompasses the compact emission detected with the $0.3''$ VLA A-configuration beam. Radio variability is also not strongly supported by the available observations: the 2010 and 2018 GHz flux densities are consistent after accounting for their frequency difference, and the three VLASS epochs show no
significant variability. The $\sim3\,\mathrm{GHz}$ VLASS measurement is consistent, within its uncertainty, with the extrapolation of the optically thin $6\,\mathrm{GHz}$ spectrum, while a steeper spectral index would predict an even higher 144\,MHz flux density.
Together, these results favour spectral curvature or absorption below the optically thin GHz band, while leaving the turnover frequency uncertain. Simultaneous broadband data are needed to distinguish absorption models and possible variability.

\paragraph{J1717.}
J171741.74+380752 is the only \citet{2026arXiv260521574L} gold-sample source in the detected list. The VLASS inverse-variance mean is $0.29\pm0.07\,\mathrm{mJy}$ ($4.0\sigma$) and LoTSS gives $0.20\pm0.10\,\mathrm{mJy}$ ($2.0\sigma$) at the optical position. It is the only known Balmer-absorption LRD with a candidate compact GHz counterpart, and a single deeper 3\,GHz pointing would confirm or exclude its radio detection.
 
\medskip
\noindent At one frequency the detected sources span more than two orders of magnitude. Against $L_{144}=3.6\times10^{31}\,\mathrm{erg\,s^{-1}\,Hz^{-1}}$ for J2048 stands $L_{144}<2.9\times10^{29}\,\mathrm{erg\,s^{-1}\,Hz^{-1}}$ ($3\sigma$) for J1047, a factor of 125 between two objects that have the same category label, same redshift range and the secure radio measurements. Restricted to the five LoTSS detections, the span runs from $1.8\times10^{30}$ to $3.6\times10^{31}\,\mathrm{erg\,s^{-1}\,Hz^{-1}}$ (Fig.~\ref{fig:lz}, left panel).

\subsection{Radio spectrum}
\label{sec:spec}

\begin{figure*}
\centering
\includegraphics[width=0.9\textwidth]{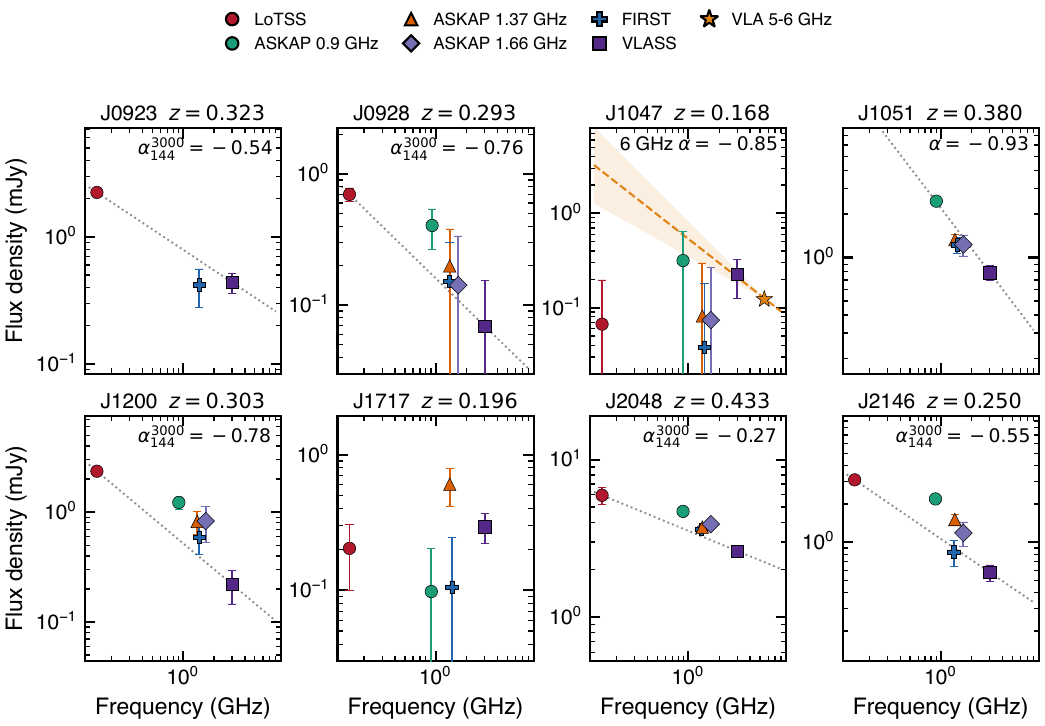}
\caption{Radio spectra of the eight detected sources. Points are $S\pm\sigma$ forced photometry from LoTSS, FIRST, ASKAP and VLASS, and NVSS, TGSS and GLEAM are omitted. Dotted lines, where drawn, are the LoTSS--VLASS two-point spectral index (Sect.~\ref{sec:spec}). For J1047 the dashed line and shaded band are the 6\,GHz power law of \citet{2026A&A...707L..17R} with $\alpha=-0.85\pm0.24$, and its 144\,MHz point is the LoTSS amplitude measured at the optical position.}
\label{fig:sed}
\end{figure*}

As the radio surveys are not beam-matched, we adopt the two-point LoTSS--VLASS spectral index $\alpha_{144}^{3000}$ as the spectral descriptor and use the intermediate bands to constrain the extended contribution (Fig.~\ref{fig:sed}). For J2146 the LoTSS--VLASS track predicts $0.89\,\mathrm{mJy}$ at 1.4\,GHz and FIRST, at $5.4''$, measures $0.84\pm0.20\,\mathrm{mJy}$, while the ASKAP 0.9\,GHz mean, on beams of $13''$--$17''$, exceeds the same track by $\approx1\,\mathrm{mJy}$. The excess therefore sits on angular scales larger than the FIRST beam, or in the relative flux-density scales of the ASKAP products, and in either case it leaves $\alpha_{144}^{3000}$ untouched. No single power law passes through all the bands of J2146 and J2048, the mismatch being carried by the large-beam points, while the two-point spectral index shifts by only a few hundredths. 

\textit{Flat.} J2048 has $\alpha_{144}^{3000}=-0.27\pm0.04$, and this flat spectral shape is consistent across the available frequency pairs. Between LoTSS (a $9''$ beam) and FIRST (a $5.4''$ beam) the spectral index is $-0.22$. Between FIRST and VLASS (a $2.5''$ beam) it is $-0.42$. Star-formation-dominated synchrotron emission at these frequencies has $\alpha\simeq-0.7$ to $-0.8$, so the low-frequency spectrum of J2048 is flatter than its ULIRG luminosity would lead one to expect. Either a flat-spectrum compact component contributes at metre wavelengths, or free--free absorption in the starburst can flatten the spectrum below $\sim1\,\mathrm{GHz}$. A VLBI brightness temperature will help to separate the two scenarios.

\textit{Intermediate.} J0923 has $\alpha_{144}^{3000}=-0.54\pm0.06$ with FIRST and VLASS in agreement, and J2146 has $-0.55\pm0.05$ with the extended contribution quantified above. These are ordinary optically thin synchrotron spectra.

\textit{Steep.} J1200 has $\alpha_{144}^{3000}=-0.78\pm0.11$. ASKAP-low, on a $25''$ beam, is fainter than LoTSS on $6''$, and FIRST at $5.4''$ continues the same decline. J0928 is the extreme case, a secure 144\,MHz detection with no 3\,GHz counterpart. The spectral index  $\alpha_{144}^{3000}=-0.76\pm0.40$ of J0928 depends on a low-significance VLASS amplitude and is only an illustrative spectral descriptor. J1051, which has no LoTSS coverage, is steep between 0.9 and 3\,GHz, $\alpha_{0.9}^{3}\simeq-0.93$.

\begin{figure}
\centering
\includegraphics[width=\hsize]{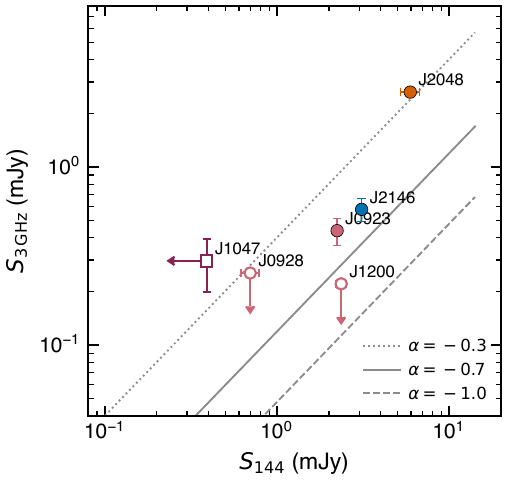}
\caption{144\,MHz flux density versus the mean VLASS 3\,GHz flux density for sources with LoTSS coverage. Filled circles are detections in both bands. Open symbols with arrows denote sources with an upper limit in one band. Reference lines show single power-law spectra, $S_\nu \propto \nu^\alpha$, for the labelled spectral indices. J1047 lies to the left of the flattest reference line, reflecting the absence of a 144\,MHz counterpart to its published 6\,GHz source.}
\label{fig:plane}
\end{figure}

\textit{Low-frequency suppressed.} J1047 is the one source whose GHz power law overpredicts the metre-wave measurement by a large factor. In the $S_{144}$--$S_{3}$ plane it falls to the left of the $\alpha=-0.3$ track (Fig.~\ref{fig:plane}), while the other sources with LoTSS coverage lie along tracks between $\alpha\simeq-0.3$ (J2048) and $\alpha\simeq-0.8$ (J1200).

Two observational properties characterize its radio spectra. The first is the radio spectral index, which runs from $-0.27$ to $-0.78$ between 144\,MHz and 3\,GHz and reaches $\simeq-0.93$ between 0.9 and 3\,GHz in J1051, the one detected source without a 144\,MHz measurement. The second is low-frequency suppression, unambiguous in J1047 and possibly already at work in J2048. The present data do not uniquely identify free–free absorption, synchrotron self-absorption, or a combination. Opacity is also the natural link between the optical and the radio properties. The studies that define the LRD class infer large gas columns on sub-parsec scales \citep{Rusakov2025,InayoshiHo2026}, and the radio spectra differ from source to source most clearly in the degree of low-frequency suppression.

\subsection{Persistence}
\label{sec:var}

The spectra in Sect.~\ref{sec:spec} combine  non-simultaneous archival measurements spanning several decades. Multi-epoch detections establish persistent radio components in some sources, but do not demonstrate ten-per-cent stability over the full interval. We therefore interpret the assembled spectra as representative archival states, with possible variability and angular-resolution differences retained as uncertainties.

\begin{figure*}
\centering
\includegraphics[width=\textwidth]{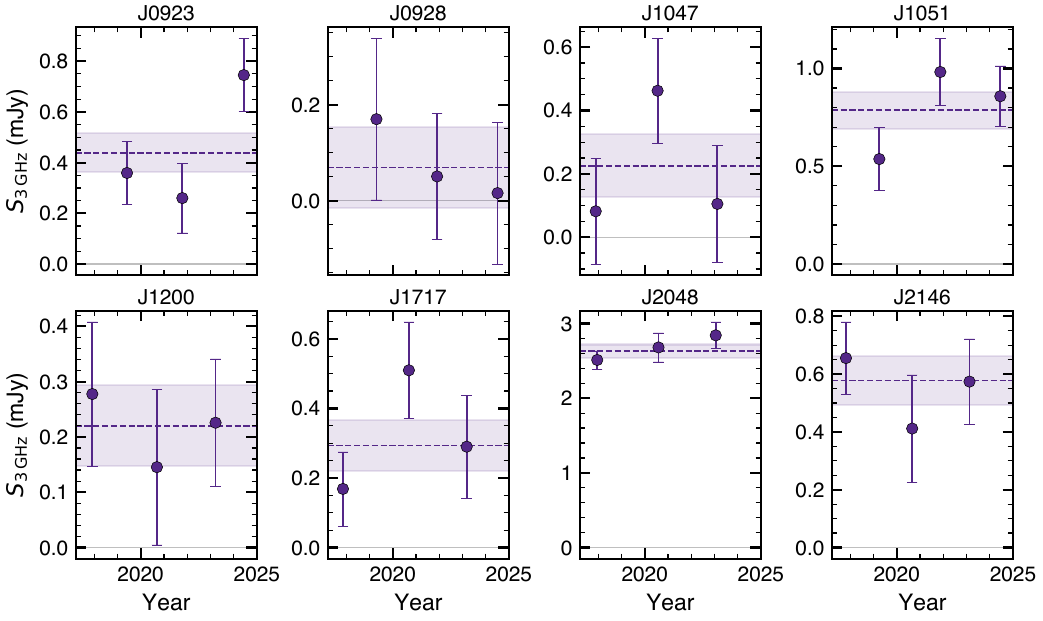}
\caption{VLASS 3\,GHz flux densities at the optical position. The plotted values represent forced-photometry measurements only and should not be interpreted as individual detections. The dashed line and band are the inverse-variance mean and its statistical uncertainty. The shaded band contains statistical errors only. Flux-density-scale systematics and the limited significance of epoch differences are discussed in Sect.~\ref{sec:var}).}
\label{fig:vlasslc}
\end{figure*}

VLASS supplies three epochs at the same frequency and comparable resolution  (Fig.~\ref{fig:vlasslc}). J0923 shows the largest change of the eight: epoch values of $0.36\pm0.14$, $0.26\pm0.13$ and $0.75\pm0.17\,\mathrm{mJy}$ give a peak-to-peak spread of $0.49\,\mathrm{mJy}$ about a mean of $0.44\pm0.08\,\mathrm{mJy}$. This spread exceeds 20 per cent of its mean and cannot be described as a calibration-scale excursion alone. The longer-term brightening discussed in Paper I must nevertheless be retained when interpreting its archival spectrum. J2048, the brightest VLASS source, stays at $2.63\,\mathrm{mJy}$ with a 12\,per cent peak-to-peak range, well inside the same flux-density-scale budget \citep{NRAO2026}.

ASKAP-low light curves exist for J1051, J1200, J2048 and J2146 (Fig.~\ref{fig:askaplc}; Appendix~\ref{app:lc}). J2048 has 33 epochs between 2019 and 2026. The rms scatter is 5\,per cent, on restoring beams that range from $13''$ to $25''$. J1051 has four epochs and a peak-to-peak change of 24\,per cent, again with mixed beams. Beam and flux-density-scale differences between RACS/VAST products complicate tests of intrinsic variability. These measurements establish multi-year persistence of J2048 and J1051. Archival photometry spanning three decades can therefore be assembled into a single spectral energy distribution, and the 144\,MHz deficit of J1047 cannot be attributed to a source caught in different states at different epochs.

\section{Discussion}
\label{sec:disc}

\subsection{One optical class, several radio states}
\label{sec:nopheno}
 
The radio-detected sample comprises three V-shaped Green Peas, one ULIRG analogue, two DESI analogues, one local SDSS LRD with a published compact radio core, and one gold LRD with Balmer absorption. At 144\,MHz, the measurements span at least a factor of 125 in luminosity, from $L_{144}=3.6\times10^{31}\,\mathrm{erg\,s^{-1}\,Hz^{-1}}$ for J2048 to the $3\sigma$ limit of $L_{144}<2.9\times10^{29}\,\mathrm{erg\,s^{-1}\,Hz^{-1}}$ for J1047. Their radio spectra are similarly diverse, ranging from relatively flat ($\alpha\simeq-0.3$) to steep ($\alpha\simeq-0.8$), with J1047 showing a low-frequency turnover above 144\,MHz. Neither radio luminosity nor spectral shape shows a simple correspondence with the optical subclasses.
 
The apparent differences in LoTSS detection rates are strongly affected by survey depth, and quantifying this selection effect is an important part of the census. Among the 35 targets with LoTSS coverage, three of the seven Green Peas are detected, compared with two of the other 28 targets. Restricting the comparison to the 23 targets whose $3\sigma$ luminosity limits fall below the luminosity of J0928 leaves three detections among six Green Peas and two among 17 other sources. The Green Peas lie mainly at $z\simeq0.3$, where LoTSS reaches $L_{144}\sim10^{30}\,\mathrm{erg\,s^{-1}\,Hz^{-1}}$, whereas the gold sample lies mainly at $z\simeq0.6$, where the same luminosity cannot be reached. Differences in luminosity sensitivity therefore contribute substantially to the apparent excess of LoTSS detections among the Green Peas.

Balmer absorption shows a similar dependence on redshift and survey depth. None of the five LoTSS detections is flagged for Balmer absorption, compared with 15 of the 30 non-detections, and the flagged sources are concentrated mainly in the higher-redshift gold sample. The three Balmer-flagged sources at $z<0.23$, J1022, J1025, and J1717, have 144\,MHz limits below the luminosities of the detected Green Peas; J1022 and J1025 nevertheless remain undetected at this depth. J1047, the only source with a clear low-frequency deficit, has no Balmer-absorption flag in the parent catalogue. Any apparent association between Balmer absorption and metre-wave weakness therefore cannot presently be separated from the redshift and sensitivity differences described above.

\subsection{Detection fractions and how the census should be compared}
\label{sec:fraction}
 
\citet{2026arXiv260816200R} report a 29 per cent VLASS association rate (14 of 48 objects) using a catalogue-based positional match. Applying the same S/N$\ge5$ threshold to our forced-photometry pipeline gives four VLASS detections in 44 targets (J0923, J1051, J2048 and J2146). The two rates are not directly comparable: forced photometry measures the beam-scale flux density centred on the optical nucleus, whereas a catalogue-based association accepts any source within a search radius and can therefore include offset counterparts, extended host emission, and positive noise excursions. Our lower S/N$\ge3$ census adds J1200 and J1717, with J1200 independently detected at 16$\sigma$ in LoTSS and J1717 supported by its multi-epoch VLASS mean, bringing the VLASS count to six.

Including the LoTSS detections and the published VLA core of J1047 brings the total number of radio-detected sources to eight. A single detection fraction for the full sample, however, has limited physical meaning. Nine targets lack usable LoTSS coverage, the $z\gtrsim0.5$ gold sample is largely limited by survey sensitivity, and J2048 is substantially more radio luminous than the rest of the sample. Any overall detection fraction would therefore combine the effects of redshift, survey depth, and parent-sample selection rather than isolate their individual contributions.

Figure~\ref{fig:lz} shows the luminosity measurements and limits source by source. At $z\lesssim0.35$ and $L_{144}\gtrsim2\times10^{30}\,\mathrm{erg\,s^{-1}\,Hz^{-1}}$, the sample contains both detections and comparably deep non-detections, demonstrating genuine source-to-source diversity at matched sensitivity. At $z\gtrsim0.5$, all sources are represented by upper limits, and these limits remain above the luminosity of the faintest low-redshift detection, J0928. The low-redshift sample is therefore inconsistent with a single characteristic radio luminosity for the LRD population, while the detections of J2048, J0923, J2146, and J1200 rule out uniform radio silence. J1047 provides a complementary selection effect: its 144\,MHz upper limit lies a factor of seven below the extrapolation of its GHz spectrum. A metre-wave-only survey could therefore miss LRDs in this low-frequency-suppressed spectral state, making a LoTSS-selected local sample incomplete with respect to radio spectral diversity.
 
\subsection{What the local sample implies at $z\gtrsim6$}
\label{sec:highz}

\begin{figure}
\centering
\includegraphics[width=\hsize]{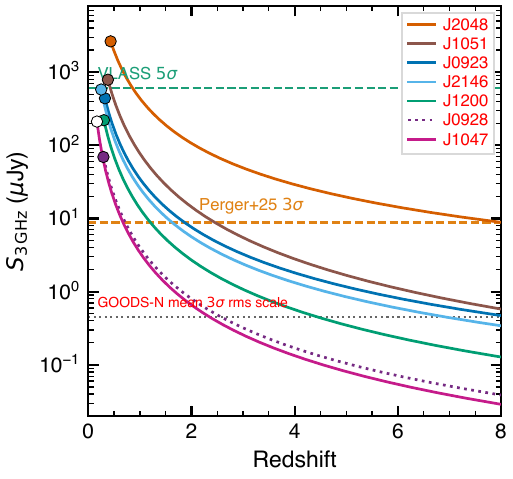}
\caption{Predicted 3\,GHz flux density for seven local radio-detected LRDs redshifted to higher $z$, computed using each source's measured two-point spectral index and the luminosity-distance scaling $S(z)=S_0(D_{\rm L,0}/D_{\rm L})^2[(1+z)/(1+z_0)]^{1+\alpha}$. J0928 is shown dotted because its slope is poorly constrained. The upper dashed line is the typical VLASS Quick Look 5$\sigma$ depth (0.60 mJy), and the lower dashed line is the VLASS 3 GHz 3$\sigma$ mean-stack limit of \citet{2025A&A...693L...2P} (8.8 $\mu$Jy)  Filled circles mark the native detection redshifts. The dotted horizontal line marks $3\sigma \sim 0.45 \mu$Jy, the noise level inferred for the GOODS-N broad-line AGN of \citet{2026A&A...706A.372M}.}
\label{fig:highz}
\end{figure}

\begin{figure}
\centering
\includegraphics[width=\hsize]{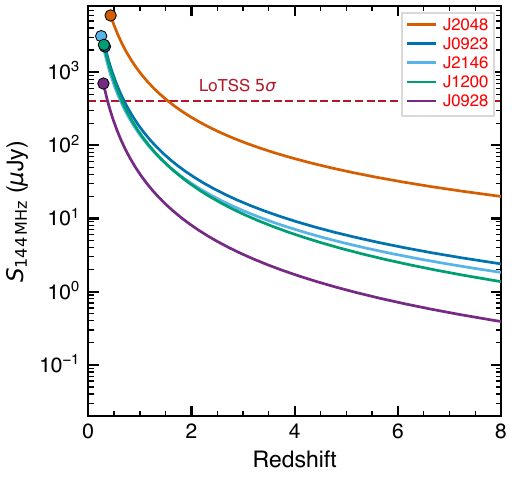}
\caption{Predicted 144\,MHz flux density for the five LoTSS-detected sources (J2048, J0923, J2146, J1200 and J0928) redshifted to higher $z$, using the same luminosity-distance scaling as Fig.~\ref{fig:highz}. J1047 is omitted because it is not detected at 144\,MHz. Filled circles mark the native detection redshifts. The dashed line is the typical LoTSS $5\sigma$ depth ($0.40\,\mathrm{mJy}$).}
\label{fig:highz144}
\end{figure}

Figures~\ref{fig:highz} and~\ref{fig:highz144} give conditional predictions obtained by moving the already visible radio components of local analogues to higher redshift while holding their escaping rest-frame spectra fixed.
For a power law, $S_\nu \propto \nu^\alpha$, the flux-density scaling is $S(z)=S_0(D_{\rm L,0}/D_{\rm L})^2[(1+z)/(1+z_0)]^{1+\alpha}$. At $z=6$, the observed-frame 3\,GHz flux densities are approximately $14\,\mu$Jy (J2048), $1.1\,\mu$Jy (J1051), $0.8\,\mu$Jy (J0923), $0.6\,\mu$Jy (J2146), $0.24\,\mu$Jy (J1200), and $0.07\,\mu$Jy (J0928). J1047 gives $\sim0.05\,\mu$Jy when anchored to its secure 6 GHz core and the in-band optically thin index. J1051 uses $\alpha_{0.9}^{3}=-0.93$. J0928's track is illustrative because its 3\,GHz amplitude does not securely determine a slope. J1717 is omitted because its spectrum is insufficiently constrained. At observed-frame 144\,MHz, the five LoTSS-detected sources reach $32$, $4.1$, $3.2$, $2.5$ and $0.7\,\mu$Jy (J2048, J0923, J2146, J1200 and J0928); J1047 is excluded because it is not detected at 144\,MHz. Moving to $z=7$ reduces all values by roughly 25 per cent.

Observed 3\,GHz at $z=6$ samples rest-frame 21\,GHz, beyond the frequency range used to measure most of the local two-point indices. Spectral curvature can therefore change these values, particularly for the flat low-frequency spectrum of J2048 and the uncertain slope of J0928. Evolution in luminosity, radio-component incidence, absorbing geometry, or magnetic field is not included. Inverse-Compton losses against the cosmic microwave background may further affect extended components with weak magnetic fields. 

If the typical high-redshift LRD hosted a source similar to J0923 or J2146, the stacked 3\,GHz mean at $z\sim6$ would be $\sim0.7\,\mu$Jy, below the wide-area VLASS mean-stack limit but not below all existing stacking sensitivity. In GOODS-N, \citet{2026A&A...706A.372M} obtain a 3 GHz mean-stack rms of 0.15 $\mu$Jy, corresponding to a noise-defined 3$\sigma$ scale of 0.45 $\mu$Jy. Such a deep stack can test substantial contributions from these templates. A direct population comparison must match the broad-line AGN versus LRD selection, the redshift distribution, and the stacking statistic; a confidence upper bound must also account for the measured stacked amplitude.

For a population in which a fraction $f$ of objects are J2048 analogues and the rest are of the common local type, the predicted stacked mean is $\bar{S}_{3}\simeq f\times14\,\mu{\rm Jy}+(1-f)\times0.7\,\mu{\rm Jy}$, giving $f\lesssim(\bar{S}_{3}-0.7\,\mu{\rm Jy})/13\,\mu{\rm Jy}$. The current VLASS and FIRST upper limit already rules out a J2048-dominated high-redshift population; a microjansky-level stacked mean would restrict J2048-like sources to a few per cent of LRDs at $z\sim6$.

Two observational consequences follow. First, current observations probe only the bright tail of the population, since the typical local radio state falls below existing stacking sensitivity at $z\gtrsim5$. Second, if J1047-like sources, GHz-bright but metre-wave suppressed, exist at high redshift, then observed-frame 144\,MHz surveys at $z\sim6$ sample rest-frame $\sim1\,$GHz on the optically thick side of the spectral peak, systematically missing this population. The relevant search band shifts to 3--10\,GHz, where SKA-mid and ngVLA will reach the sub-$\mu$Jy depths required. Both points depend only on the shape of the local radio luminosity distribution, regardless of whether any individual high-redshift source turns out to be a J1047 analogue.
 
\section{Summary}
\label{sec:sum}
 
We have measured the radio flux density at the optical position of the 44 published low-redshift LRDs from 144\,MHz to 6\,GHz. Eight sources are radio-detected.  The main results are summarized as below.
 
\begin{enumerate}
\item \emph{The 144\,MHz constraints span more than two orders of magnitude.} The five LoTSS detections alone span a factor of 20 in luminosity, from $L_{144}=1.8\times10^{30}$ to $3.6\times10^{31}\,\mathrm{erg\,s^{-1}\,Hz^{-1}}$. The $3\sigma$ upper limit of $L_{144}<2.9\times10^{29}\,\mathrm{erg\,s^{-1}\,Hz^{-1}}$ for the GHz-detected source J1047 extends the constrained range to at least a factor of 125.
 
\item \emph{The spectra differ primarily in low-frequency suppression.} The spectral indices run from $-0.27\pm0.04$ (J2048) to $-0.78\pm0.11$ (J1200), and J1047 shows low-frequency suppression with an uncertain turnover frequency. Its 3\,GHz flux density sits on the published 6\,GHz power law while its 144\,MHz flux density lies a factor of seven below it. Low-frequency suppression is the primary spectral discriminant among the detected sources.
 
\item \emph{Counts track survey depth and redshift.} Redshift, survey depth, and parent-sample selection are closely coupled across this sample. The apparent excess of LoTSS detections among the Green Peas is substantially reduced once the luminosity limits are matched, indicating that survey depth accounts for much of the difference. Likewise, the discrepancy between our detection count and published association fractions largely reflects differences in counterpart-identification methods rather than necessarily intrinsic differences among the sources.
\end{enumerate}

 The local sources supply empirical radio templates for little red dots at high redshift. Placed at $z=6$ with their escaping spectra unchanged, J0923 and J2146 would reach approximately 0.6 to 0.8\,$\mu$Jy at an observed frequency of 3\,GHz, a level that deep stacks of suitably matched samples can probe. Curved J1047-like spectra are best observed above their redshifted turnover, where the required sensitivity is set by flux densities near 0.05\,$\mu$Jy at GHz frequencies. Converting these templates into a population forecast requires a joint treatment of selection, redshift distribution, spectral diversity, the incidence of visible radio components and the non-detections. The  catalogue of 44 targets in this study provides a common reference for such comparisons and a ready basis for targeted VLA, VLBI and ALMA follow-up.

\begin{acknowledgements}
AW is supported by the National Natural Science Foundation of China  (Grant Number of 12503018).  This work uses archival images from LoTSS, VLASS, ASKAP, FIRST and NVSS, and the published VLA flux-density measurements of J1047. Numerical analysis used NumPy, SciPy, pandas, Matplotlib and Astropy \citep{2022ApJ...935..167A}. 
\end{acknowledgements}

\bibliographystyle{aa}
\bibliography{ref}

\clearpage
\onecolumn
\appendix

\section{Forced photometry of the full sample}
\label{app:tab}

Table~\ref{tab:full} reports the optical-position forced amplitudes used in this paper. LoTSS entries are blank when there is no valid local image. Errors are statistical map-noise terms and do not include survey flux-scale systematics. 

{\footnotesize
\begin{longtable}{llcccccc}
\caption{Signed forced photometry for the 44-object literature union. Flux densities and uncertainties are given in mJy. Empty LoTSS entries are missing coverage (no cutout, mosaic hole, or south of the survey). Errors are statistical map-noise terms and do not include survey flux-density-scale systematics}.\\
\label{tab:full}\\
\hline\hline
Name & Parent & $z$ & $S^{\rm LoTSS}_{144 \rm MHz}$ & $\sigma^{\rm LoTSS}_{144 \rm MHz}$ & $S^{\rm VLASS}_{3 \rm GHz}$ & $\sigma^{\rm VLASS}_{3 \rm GHz}$ & Note \\
\hline
\endfirsthead
\hline\hline
Name & Parent & $z$ & $S^{\rm LoTSS}_{144 \rm MHz}$ & $\sigma^{\rm LoTSS}_{144 \rm MHz}$ & $S^{\rm VLASS}_{3 \rm GHz}$ & $\sigma^{\rm VLASS}_{3 \rm GHz}$ & Note \\
\hline
\endhead
\hline
\endfoot
J000927+081109 & Lin+26 silver & 0.3720 & -0.071 & 0.096 & -0.092 & 0.098 &  \\
J003515+084859 & Lin+25 BLGP & 0.2420 & 0.084 & 0.103 & 0.017 & 0.090 &  \\
J012930+062843 & Lin+26 gold & 0.2467 & 0.181 & 0.100 & -0.040 & 0.084 &  \\
J024337+034915 & Lin+26 silver & 0.4584 & -0.121 & 0.112 & 0.039 & 0.089 &  \\
J071635+543321 & Park+26 & 0.2908 & 0.057 & 0.052 & -0.035 & 0.070 &  \\
J082606$-$010001 & Lin+26 gold & 0.6273 & 0.323 & 0.339 & 0.036 & 0.102 &  \\
J082921+131237 & Lin+26 gold & 0.3986 & 0.043 & 0.123 & -0.064 & 0.078 &  \\
J085058+303053 & Lin+25 BLGP & 0.2800 & 0.030 & 0.087 & 0.023 & 0.082 &  \\
J085244$-$041150 & Ding+26 & 0.3700 & \ldots & \ldots & -0.004 & 0.097 &  \\
J092359+415736 & Lin+25 BLGP & 0.3230 & 2.233 & 0.104 & 0.439 & 0.077 & detected \\
J092834+292136 & Lin+25 BLGP & 0.2930 & 0.698 & 0.084 & 0.069 & 0.085 & detected \\
J092904$-$025158 & Ding+26 & 0.3400 & \ldots & \ldots & -0.095 & 0.099 &  \\
J094411$-$024908 & Lin+26 gold & 0.6623 & \ldots & \ldots & 0.120 & 0.104 &  \\
J101742+311459 & Lin+26 gold & 0.6706 & -0.119 & 0.078 & 0.024 & 0.079 &  \\
J102208+084156 & Lin+26a & 0.2227 & -0.218 & 0.080 & 0.064 & 0.089 &  \\
J102530+140207 & Lin+26a & 0.1007 & 0.068 & 0.097 & -0.101 & 0.073 & ambig. \\
J102553+502843 & Lin+26 gold & 0.8824 & 0.047 & 0.055 & -0.024 & 0.066 &  \\
J104242+372147 & Lin+26 gold & 0.6080 & -0.067 & 0.058 & 0.016 & 0.079 &  \\
J104755+073951 & Lin+26a & 0.1682 & 0.067 & 0.130 & 0.226 & 0.099 & detected \\
J105157$-$062625 & Ding+26 & 0.3800 & \ldots & \ldots & 0.786 & 0.094 & detected \\
J105620+275415 & Lin+26 silver & 0.4617 & 0.039 & 0.081 & -0.037 & 0.074 &  \\
J105900+314951 & Lin+26 silver & 0.4990 & -0.025 & 0.068 & 0.118 & 0.075 &  \\
J111943+021911 & Lin+26 silver & 0.4682 & -0.196 & 0.197 & 0.173 & 0.101 &  \\
J113734+552028 & Lin+26 silver & 0.4358 & -0.020 & 0.049 & 0.103 & 0.084 &  \\
J115438+065025 & Lin+25 BLGP & 0.2690 & -0.056 & 0.579 & 0.112 & 0.089 &  \\
J120052+331238 & Lin+25 BLGP & 0.3030 & 2.358 & 0.145 & 0.220 & 0.074 & detected \\
J122245+360218 & Lin+25 BLGP & 0.3010 & 0.048 & 0.068 & -0.028 & 0.081 &  \\
J132137$-$021417 & Lin+26 gold & 0.2244 & \ldots & \ldots & 0.008 & 0.123 &  \\
J134317+393418 & Lin+26 silver & 0.2933 & -0.039 & 0.070 & 0.098 & 0.078 &  \\
J142337+520216 & Lin+26 gold & 0.6236 & 0.067 & 0.060 & 0.055 & 0.100 &  \\
J150252+025027 & Lin+26 gold & 0.2906 & 0.017 & 0.188 & -0.099 & 0.090 &  \\
J161102+091728 & Lin+26 gold & 0.6952 & 0.278 & 0.126 & -0.044 & 0.086 &  \\
J162032+314817 & Lin+26 gold & 0.4099 & 0.005 & 0.074 & 0.092 & 0.074 &  \\
J164102+070806 & Lin+26 gold & 0.5351 & 0.163 & 0.235 & -0.055 & 0.093 &  \\
J164226+042632 & Lin+26 gold & 0.7291 & \ldots & \ldots & -0.059 & 0.093 &  \\
J164637+142648 & Lin+26 gold & 0.7071 & -0.065 & 0.066 & 0.073 & 0.074 &  \\
J165450+033741 & Lin+26 gold & 0.6408 & \ldots & \ldots & 0.033 & 0.096 &  \\
J171741+380752 & Lin+26 gold & 0.1959 & 0.204 & 0.104 & 0.294 & 0.073 & detected \\
J190954+583112 & Lin+26 silver & 0.4273 & 0.078 & 0.080 & 0.020 & 0.079 &  \\
J204837$-$002437 & Chen+26 & 0.4330 & 5.941 & 0.749 & 2.630 & 0.089 & detected \\
J212725$-$044808 & Lin+26 gold & 0.5842 & \ldots & \ldots & -0.031 & 0.084 &  \\
J214655+102813 & Ding+26 & 0.2500 & 3.098 & 0.177 & 0.578 & 0.085 & detected \\
J225535+154216 & Lin+26 gold & 0.4273 & -0.018 & 0.087 & -0.011 & 0.141 &  \\
J235751$-$112849 & Ding+26 & 0.3000 & \ldots & \ldots & 0.358 & 0.321 &  \\
\hline
\end{longtable}
}

\section{Multi-frequency photometry of the radio-detected sources}
\label{app:epochs}

Table~\ref{tab:allepochs} lists every individual-epoch forced-photometry measurement for the eight radio-detected sources, ordered by source and then by frequency. All epochs are retained separately so that variability, coverage gaps, and beam-size differences are directly visible. The per-band inverse-variance means in Table~\ref{tab:eight} are derived from these entries. 
ASKAP measurements are included for completeness and do not enter $\alpha_{144}^{3000}$. 
Observation dates are taken from the image FITS headers. Surveys for which no per-observation date is recorded in the header are indicated by dashes.

{\footnotesize
\begin{longtable}{llclll}
\caption{Individual-epoch forced photometry of the eight radio-detected sources. Values are signed amplitudes $S\pm\sigma$ in mJy fitted at the optical position, from the same release as Table~\ref{tab:full}. Errors are statistical and exclude flux-density-scale systematics. Beams are major $\times$ minor axes. Missing observation dates are shown as dashes. VLA values are published measurements and their dates are epoch years.}\label{tab:allepochs}\\
\hline\hline
Source & Survey & Date & $\nu$ (GHz) & Beam ($''$) & $S\pm\sigma$ (mJy) \\
\hline
\endfirsthead
\hline\hline
Source & Survey & Date & $\nu$ (GHz) & Beam ($''$) & $S\pm\sigma$ (mJy) \\
\hline
\endhead
\hline
\endfoot
J0923 & LoTSS & --- & 0.1440 & 6.0$\times$6.0 & $2.233\pm0.104$ \\
J0923 & ASKAP & 2022-03-18 & 0.8875 & 45.5$\times$12.2 & $1.283\pm0.334$ \\
J0923 & ASKAP & 2023-12-20 & 0.9435 & 36.4$\times$11.2 & $1.435\pm0.190$ \\
J0923 & ASKAP & 2023-12-21 & 0.9435 & 36.2$\times$11.3 & $1.088\pm0.174$ \\
J0923 & ASKAP & 2021-01-01 & 1.3675 & 39.3$\times$12.0 & $1.275\pm0.219$ \\
J0923 & ASKAP & 2024-10-06 & 1.3675 & 28.9$\times$8.7 & $0.658\pm0.353$ \\
J0923 & ASKAP & 2024-12-13 & 1.3675 & 26.9$\times$7.9 & $0.984\pm0.178$ \\
J0923 & FIRST & 1994-09-04 & 1.4351 & 5.4$\times$5.4 & $0.420\pm0.141$ \\
J0923 & ASKAP & 2021-12-24 & 1.6555 & 38.6$\times$7.9 & $0.971\pm0.226$ \\
J0923 & VLASS & 2019-05-31 & 3.0000 & 2.7$\times$2.2 & $0.359\pm0.124$ \\
J0923 & VLASS & 2021-10-17 & 3.0000 & 2.8$\times$2.2 & $0.260\pm0.137$ \\
J0923 & VLASS & 2024-06-09 & 3.0000 & 2.7$\times$2.1 & $0.745\pm0.144$ \\
J0928 & LoTSS & --- & 0.1440 & 6.0$\times$6.0 & $0.698\pm0.084$ \\
J0928 & ASKAP & 2019-04-21 & 0.8875 & 22.9$\times$11.5 & $0.439\pm0.248$ \\
J0928 & ASKAP & 2022-03-23 & 0.8875 & 26.2$\times$12.1 & $0.490\pm0.237$ \\
J0928 & ASKAP & 2023-12-20 & 0.9435 & 25.4$\times$11.1 & $-1.144\pm0.960$ \\
J0928 & ASKAP & 2023-12-21 & 0.9435 & 25.6$\times$11.3 & $1.016\pm0.783$ \\
J0928 & ASKAP & 2023-12-23 & 0.9435 & 21.8$\times$11.4 & $0.317\pm0.251$ \\
J0928 & FIRST & 1993-05-04 & 1.3649 & 5.4$\times$5.4 & $0.152\pm0.149$ \\
J0928 & ASKAP & 2021-01-03 & 1.3675 & 17.5$\times$8.2 & $0.199\pm0.182$ \\
J0928 & ASKAP & 2022-01-01 & 1.6555 & 12.4$\times$6.3 & $0.142\pm0.195$ \\
J0928 & VLASS & 2019-04-20 & 3.0000 & 3.0$\times$2.2 & $0.170\pm0.168$ \\
J0928 & VLASS & 2021-11-29 & 3.0000 & 2.6$\times$2.2 & $0.050\pm0.131$ \\
J0928 & VLASS & 2024-07-06 & 3.0000 & 2.9$\times$2.2 & $0.016\pm0.148$ \\
J1047 & LoTSS & --- & 0.1440 & 9.0$\times$9.0 & $0.067\pm0.130$ \\
J1047 & ASKAP & 2020-10-16 & 0.8875 & 25.0$\times$25.0 & $0.316\pm0.333$ \\
J1047 & ASKAP & 2021-01-17 & 1.3675 & 13.0$\times$10.6 & $0.082\pm0.217$ \\
J1047 & FIRST & 2000-01-28 & 1.4351 & 5.4$\times$5.4 & $0.038\pm0.142$ \\
J1047 & ASKAP & 2022-01-04 & 1.6555 & 11.3$\times$8.0 & $0.074\pm0.193$ \\
J1047 & VLASS & 2017-11-22 & 3.0000 & 2.7$\times$2.1 & $0.082\pm0.166$ \\
J1047 & VLASS & 2020-07-21 & 3.0000 & 2.7$\times$2.2 & $0.462\pm0.165$ \\
J1047 & VLASS & 2023-02-16 & 3.0000 & 2.5$\times$2.3 & $0.105\pm0.185$ \\
J1047 & VLA & 2010 & 5.0000 & 4.0 & $0.130\pm0.009$ \\
J1047 & VLA & 2018 & 6.0000 & 0.3 & $0.117\pm0.008$ \\
J1051 & ASKAP & 2020-10-16 & 0.8875 & 25.0$\times$25.0 & $2.797\pm0.265$ \\
J1051 & ASKAP & 2026-03-31 & 0.8875 & 14.0$\times$11.4 & $2.918\pm0.188$ \\
J1051 & ASKAP & 2026-05-26 & 0.8875 & 13.8$\times$11.4 & $2.566\pm0.200$ \\
J1051 & ASKAP & 2026-06-13 & 0.9435 & 15.0$\times$15.0 & $2.281\pm0.091$ \\
J1051 & ASKAP & 2021-01-17 & 1.3675 & 10.2$\times$8.5 & $1.349\pm0.158$ \\
J1051 & ASKAP & 2021-01-18 & 1.3675 & 10.2$\times$8.5 & $1.313\pm0.155$ \\
J1051 & FIRST & 2001-04 & 1.4351 & 5.4$\times$5.4 & $1.226\pm0.159$ \\
J1051 & ASKAP & 2022-01-11 & 1.6555 & 11.6$\times$8.5 & $1.231\pm0.206$ \\
J1051 & VLASS & 2019-04-02 & 3.0000 & 3.7$\times$2.7 & $0.537\pm0.161$ \\
J1051 & VLASS & 2021-11-16 & 3.0000 & 3.7$\times$2.1 & $0.982\pm0.171$ \\
J1051 & VLASS & 2024-06-16 & 3.0000 & 4.0$\times$2.2 & $0.858\pm0.156$ \\
J1200 & LoTSS & --- & 0.1440 & 6.0$\times$6.0 & $2.358\pm0.145$ \\
J1200 & ASKAP & 2019-04-22 & 0.8875 & 24.6$\times$10.8 & $1.488\pm0.289$ \\
J1200 & ASKAP & 2022-03-23 & 0.8875 & 36.7$\times$12.9 & $1.166\pm0.204$ \\
J1200 & ASKAP & 2024-01-25 & 0.9435 & 24.6$\times$11.0 & $0.930\pm0.416$ \\
J1200 & ASKAP & 2021-01-01 & 1.3675 & 19.0$\times$8.1 & $0.892\pm0.280$ \\
J1200 & ASKAP & 2024-10-20 & 1.3675 & 18.9$\times$8.7 & $0.760\pm0.271$ \\
J1200 & FIRST & 1999 & 1.4351 & 5.4$\times$5.4 & $0.592\pm0.177$ \\
J1200 & ASKAP & 2021-12-29 & 1.6555 & 24.5$\times$7.9 & $0.829\pm0.295$ \\
J1200 & VLASS & 2017-12-02 & 3.0000 & 3.0$\times$2.3 & $0.278\pm0.130$ \\
J1200 & VLASS & 2020-09-12 & 3.0000 & 3.2$\times$2.4 & $0.145\pm0.141$ \\
J1200 & VLASS & 2023-03-23 & 3.0000 & 2.5$\times$2.3 & $0.226\pm0.115$ \\
J1717 & LoTSS & --- & 0.1440 & 6.0$\times$6.0 & $0.204\pm0.104$ \\
J1717 & ASKAP & 2023-06-30 & 0.8425 & 43.6$\times$13.4 & $0.483\pm0.321$ \\
J1717 & ASKAP & 2022-03-23 & 0.8875 & 33.6$\times$12.1 & $0.299\pm0.192$ \\
J1717 & ASKAP & 2023-12-24 & 0.9435 & 30.6$\times$11.4 & $0.236\pm0.221$ \\
J1717 & ASKAP & 2024-04-06 & 0.9435 & 29.9$\times$11.2 & $-0.265\pm0.173$ \\
J1717 & ASKAP & 2024-11-01 & 1.3675 & 22.1$\times$8.0 & $0.606\pm0.192$ \\
J1717 & FIRST & 1994-08-05 & 1.4351 & 5.4$\times$5.4 & $0.106\pm0.141$ \\
J1717 & VLASS & 2017-11-10 & 3.0000 & 2.5$\times$2.4 & $0.169\pm0.107$ \\
J1717 & VLASS & 2020-09-12 & 3.0000 & 2.6$\times$2.5 & $0.510\pm0.139$ \\
J1717 & VLASS & 2023-03-09 & 3.0000 & 2.9$\times$2.2 & $0.290\pm0.148$ \\
J2048 & LoTSS & --- & 0.1440 & 9.0$\times$9.0 & $5.941\pm0.749$ \\
J2048 & ASKAP & 2021-12-20 & 0.8555 & 14.5$\times$13.1 & $4.761\pm0.096$ \\
J2048 & ASKAP & 2024-06-13 & 0.8555 & 13.8$\times$12.7 & $4.773\pm0.089$ \\
J2048 & ASKAP & 2019-04-27 & 0.8875 & 17.6$\times$11.2 & $4.118\pm0.257$ \\
J2048 & ASKAP & 2019-08-27 & 0.8875 & 14.5$\times$13.2 & $4.898\pm0.297$ \\
J2048 & ASKAP & 2019-10-29 & 0.8875 & 13.5$\times$11.4 & $4.509\pm0.241$ \\
J2048 & ASKAP & 2019-10-30 & 0.8875 & 14.7$\times$12.1 & $5.028\pm0.269$ \\
J2048 & ASKAP & 2020-01-11 & 0.8875 & 14.4$\times$11.2 & $5.005\pm0.282$ \\
J2048 & ASKAP & 2020-01-12 & 0.8875 & 19.0$\times$11.1 & $5.184\pm0.233$ \\
J2048 & ASKAP & 2020-01-19 & 0.8875 & 14.2$\times$11.3 & $4.699\pm0.272$ \\
J2048 & ASKAP & 2020-06-19 & 0.8875 & 14.4$\times$11.8 & $4.190\pm0.210$ \\
J2048 & ASKAP & 2020-08-29 & 0.8875 & 13.6$\times$11.4 & $5.373\pm0.304$ \\
J2048 & ASKAP & 2020-10-16 & 0.8875 & 25.0$\times$25.0 & $4.663\pm0.407$ \\
J2048 & ASKAP & 2021-07-23 & 0.8875 & 14.2$\times$12.1 & $4.745\pm0.196$ \\
J2048 & ASKAP & 2021-08-21 & 0.8875 & 14.1$\times$12.1 & $4.156\pm0.226$ \\
J2048 & ASKAP & 2022-04-11 & 0.8875 & 13.4$\times$11.3 & $4.164\pm0.213$ \\
J2048 & ASKAP & 2023-07-02 & 0.8875 & 14.2$\times$12.0 & $4.727\pm0.208$ \\
J2048 & ASKAP & 2023-07-04 & 0.8875 & 15.3$\times$12.1 & $4.732\pm0.241$ \\
J2048 & ASKAP & 2023-08-29 & 0.8875 & 15.0$\times$11.9 & $4.588\pm0.198$ \\
J2048 & ASKAP & 2023-10-26 & 0.8875 & 15.3$\times$12.7 & $4.046\pm0.255$ \\
J2048 & ASKAP & 2024-04-17 & 0.8875 & 15.3$\times$12.4 & $4.181\pm0.260$ \\
J2048 & ASKAP & 2024-06-15 & 0.8875 & 15.1$\times$11.8 & $4.875\pm0.251$ \\
J2048 & ASKAP & 2024-08-12 & 0.8875 & 14.5$\times$11.5 & $4.542\pm0.230$ \\
J2048 & ASKAP & 2024-10-14 & 0.8875 & 16.6$\times$12.4 & $4.677\pm0.196$ \\
J2048 & ASKAP & 2024-12-20 & 0.8875 & 18.1$\times$13.3 & $4.456\pm0.219$ \\
J2048 & ASKAP & 2025-02-21 & 0.8875 & 17.9$\times$11.9 & $4.883\pm0.252$ \\
J2048 & ASKAP & 2025-04-22 & 0.8875 & 14.1$\times$11.7 & $4.801\pm0.248$ \\
J2048 & ASKAP & 2025-07-12 & 0.8875 & 15.4$\times$12.5 & $4.979\pm0.259$ \\
J2048 & ASKAP & 2025-09-08 & 0.8875 & 15.5$\times$13.3 & $4.672\pm0.227$ \\
J2048 & ASKAP & 2025-11-03 & 0.8875 & 14.1$\times$12.7 & $5.027\pm0.218$ \\
J2048 & ASKAP & 2026-01-21 & 0.8875 & 14.3$\times$13.0 & $5.008\pm0.265$ \\
J2048 & ASKAP & 2026-03-25 & 0.8875 & 16.2$\times$12.2 & $5.065\pm0.239$ \\
J2048 & ASKAP & 2026-05-26 & 0.8875 & 16.0$\times$12.1 & $4.642\pm0.188$ \\
J2048 & ASKAP & 2024-01-03 & 0.9435 & 13.9$\times$10.8 & $4.701\pm0.206$ \\
J2048 & FIRST & 2011-04-24 & 1.3340 & 5.4$\times$5.4 & $3.609\pm0.164$ \\
J2048 & ASKAP & 2021-01-17 & 1.3675 & 11.1$\times$9.6 & $3.950\pm0.179$ \\
J2048 & ASKAP & 2021-07-29 & 1.3675 & 9.4$\times$7.6 & $3.646\pm0.203$ \\
J2048 & ASKAP & 2021-09-21 & 1.3675 & 14.4$\times$7.7 & $3.937\pm0.215$ \\
J2048 & ASKAP & 2021-11-22 & 1.3675 & 9.5$\times$7.7 & $3.467\pm0.236$ \\
J2048 & ASKAP & 2024-12-11 & 1.3675 & 9.7$\times$7.5 & $3.484\pm0.227$ \\
J2048 & ASKAP & 2022-01-10 & 1.6555 & 11.7$\times$8.5 & $3.896\pm0.207$ \\
J2048 & VLASS & 2017-12-16 & 3.0000 & 3.6$\times$2.3 & $2.512\pm0.122$ \\
J2048 & VLASS & 2020-08-05 & 3.0000 & 2.9$\times$2.4 & $2.680\pm0.195$ \\
J2048 & VLASS & 2023-01-21 & 3.0000 & 2.7$\times$2.1 & $2.842\pm0.179$ \\
J2146 & LoTSS & --- & 0.1440 & 9.0$\times$9.0 & $3.098\pm0.177$ \\
J2146 & ASKAP & 2023-07-26 & 0.8425 & 15.9$\times$13.4 & $2.012\pm0.502$ \\
J2146 & ASKAP & 2025-12-24 & 0.8555 & 22.5$\times$12.4 & $2.455\pm0.183$ \\
J2146 & ASKAP & 2019-04-26 & 0.8875 & 21.0$\times$17.9 & $2.033\pm0.479$ \\
J2146 & ASKAP & 2020-10-17 & 0.8875 & 25.0$\times$25.0 & $2.615\pm0.655$ \\
J2146 & ASKAP & 2022-04-16 & 0.8875 & 19.5$\times$15.3 & $2.225\pm0.257$ \\
J2146 & ASKAP & 2022-10-29 & 0.8875 & 16.2$\times$13.4 & $2.012\pm0.362$ \\
J2146 & ASKAP & 2024-01-02 & 0.9435 & 13.1$\times$12.4 & $1.965\pm0.185$ \\
J2146 & FIRST & 2011-04-23 & 1.3340 & 5.4$\times$5.4 & $0.839\pm0.195$ \\
J2146 & ASKAP & 2021-01-11 & 1.3675 & 13.0$\times$10.8 & $1.387\pm0.203$ \\
J2146 & ASKAP & 2024-12-15 & 1.3675 & 9.5$\times$8.9 & $1.663\pm0.228$ \\
J2146 & ASKAP & 2022-01-04 & 1.6555 & 13.2$\times$8.7 & $1.183\pm0.262$ \\
J2146 & VLASS & 2017-10-25 & 3.0000 & 2.9$\times$2.3 & $0.655\pm0.125$ \\
J2146 & VLASS & 2020-08-31 & 3.0000 & 2.7$\times$2.5 & $0.412\pm0.186$ \\
J2146 & VLASS & 2023-02-17 & 3.0000 & 3.3$\times$2.3 & $0.574\pm0.148$ \\
\end{longtable}
}

\section{ASKAP light curves of the detected sources}
\label{app:lc}

\begin{figure*}
\centering
\includegraphics[width=\textwidth]{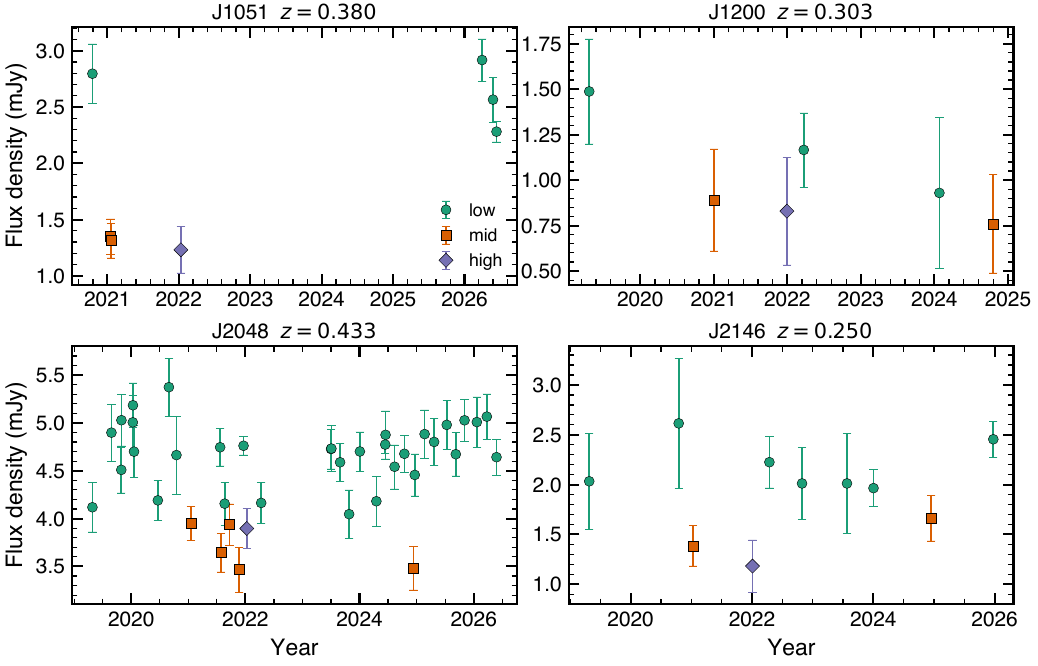}
\caption{ASKAP forced photometry for the four detected sources with useful time coverage. Green circles: low band ($\sim0.89\,\mathrm{GHz}$); orange squares: mid band (1.37\,GHz); purple diamonds: high band (1.66\,GHz). Beam and flux-density-scale differences among products are not removed.}
\label{fig:askaplc}
\end{figure*}

\end{document}